\documentclass[11pt]{article}
\usepackage[a4paper,margin=2.3cm]{geometry}
\usepackage{amssymb}
\usepackage{amsmath}
\usepackage{graphicx}
\usepackage{booktabs}
\usepackage{url}
\usepackage[hidelinks]{hyperref}

\title{A Topological--Geometrical Descriptor (TGD) for Scalable and Interpretable Modeling of Two-Dimensional Carbon Networks}

\author{R\^omulo Uriel Liberal de Oliveira$^{1}$ and Raphael Matozo Tromer$^{2,*}$\\[0.4em]
\small $^{1}$Department of Computational Engineering, University of Bras\'ilia, 70910-900, Federal District, Bras\'ilia, Brazil\\
\small $^{2}$Institute of Physics, University of Bras\'ilia, 70910-900, Federal District, Bras\'ilia, Brazil\\
\small $^{*}$Corresponding author: raphael.tromer@unb.br}
\date{}

\begin{document}
\maketitle
\begin{abstract}
We introduce a compact Topological--Geometrical Descriptor (TGD) for planar two-dimensional carbon networks. Atomic coordinates are mapped to a periodic planar graph and represented through ring populations, topological-charge statistics, projected ring geometry, and dual-graph defect adjacency. The resulting modeling vector contains 19 physically interpretable features. We evaluate TGD on 120 published planar carbon allotropes using the DFT relative energy per atom with respect to graphene as the target. Repeated five-fold cross-validation with gradient boosting yields a mean absolute error (MAE) of $0.0826\pm0.0138$ eV atom$^{-1}$ and $R^2=0.803\pm0.074$. Under the same validation protocol, TGD gives a mean MAE of $0.0828$ eV atom$^{-1}$, compared with $0.0929$ eV atom$^{-1}$ for a 203-feature Matminer representation. We further analyze 360 vacancy-defected graphene configurations spanning 12 vacancy concentrations from 2 to 24\%. When entire vacancy concentrations are held out, TGD gives a pooled MAE of $0.269$ eV vacancy$^{-1}$ and $R^2=0.861$, outperforming simple baselines based on vacancy concentration alone or non-hexagonal-ring fraction alone. Sensitivity tests show that the inferred edge sets are stable to moderate changes in the adaptive bond factor. Finally, the mean absolute topological charge is examined as a one-variable empirical surrogate for the allotrope dataset and is treated explicitly as dataset-specific rather than universal. These results show that TGD combines compact prediction with direct structural interpretation while making clear the present restriction to single-layer planar carbon networks.
\end{abstract}


\section{Introduction}
Two-dimensional carbon materials span dense graphene-like lattices, non-hexagonal defect networks, and porous $sp/sp^2$ architectures.\cite{Swiegers2000,Wang2017,Novoselov2012,Wang2019,Kumar2005,Fan2021} Their energetic and functional diversity motivates structural representations that are simultaneously compact, reproducible, and physically interpretable. General-purpose materials featurizers such as Matminer are valuable because they provide broad families of descriptors,\cite{matminer} whereas domain-specific representations can exploit known structural organization to reduce dimensionality and expose interpretable trends.

Ring statistics are a long-established description of medium-range order in network solids, with important distinctions between face rings and shortest-path definitions.\cite{Franzblau1991} In two-dimensional carbon, experimentally resolved amorphous monolayers exhibit broad distributions of five-, six-, seven-, and eight-membered rings,\cite{Toh2020} and recent atomistic work on amorphous graphene has likewise used ring statistics to quantify medium-range order.\cite{ElMachachi2022} The contribution of the present work is therefore not the use of rings by itself, but their integration with projected ring geometry, topological-charge moments, and dual-graph defect adjacency into one compact feature vector for interpretable machine learning.

Here we develop a Topological--Geometrical Descriptor (TGD) for \emph{planar} carbon networks. We first validate graph construction on graphene, phagraphene, and a porous carbon-ene-yne network. We then evaluate predictive accuracy with repeated cross-validation rather than a single train/test split, define all energy targets and units explicitly, and compare TGD with Matminer under identical validation conditions. For vacancy-defected graphene, we use group-aware validation in which complete vacancy concentrations are held out and we add two deliberately simple controls: vacancy concentration alone and non-hexagon fraction alone. These controls test whether the descriptor carries information beyond defect count. We additionally quantify sensitivity of the adaptive bond criterion and state where the current single-cutoff and planar-face assumptions may fail.

The analysis is organized to distinguish two claims. \emph{Predictive accuracy} is assessed statistically through cross-validation and explicit baselines. \emph{Physical interpretability} is assessed through the meaning of the descriptor components and through structural trends across the datasets. We avoid treating either one as evidence for the other.

\section{Methods}
\subsection{Scope and graph construction}
The present implementation assumes a single, planar, periodically repeated carbon network that can be represented by an embedded planar graph $G=(V,E)$. This scope includes the planar $sp^2$ allotropes and planar mixed $sp/sp^2$ porous networks investigated here. It does \emph{not} presently provide a unique face construction for strongly buckled sheets, curved membranes, nanotubes, multilayers, or intersecting/non-planar networks. In those cases, projection onto the $xy$ plane can alter ring areas or edge crossings, and a surface-aware or layer-resolved extension would be required.

Each structure is replicated in plane and treated with periodic boundary conditions and the minimum-image convention. Candidate neighbors are generated within $r_{\max}=3.5$~\AA, retaining at most $k_{\rm candidates}=12$ nearest candidates per atom. The shortest candidate distance is
\begin{equation}
 d_0=\min_{i\neq j}d_{ij},
\end{equation}
and the adaptive threshold is
\begin{equation}
 d_{\max}=\alpha d_0, \qquad \alpha=1.40.
\end{equation}
An edge is retained only when $d_{ij}\le d_{\max}$ and the neighbor relation is reciprocal. This construction permits both two-coordinate $sp$ and three-coordinate $sp^2$ environments; threefold coordination is not imposed. The use of a global $d_0$ is convenient but not universal: an anomalously short reconstructed bond can reduce $d_{\max}$ enough to omit longer genuine bonds. We therefore quantify sensitivity to $\alpha$ below and regard unusually broad bond-length distributions as a condition requiring explicit connectivity checks.

Faces are traced with a deterministic directed half-edge ``turn-left'' traversal of the planar embedding, a standard representation for planar maps and polyhedral surfaces.\cite{Kettner1999} Thus, TGD uses the faces of the embedded planar graph as its rings; this is distinct from a shortest-path ring set, which is another established convention for network-ring statistics.\cite{Franzblau1991} For directed edge $i\rightarrow j$, the next neighbor $k$ minimizes the positive counterclockwise angle
\begin{equation}
\theta(\mathbf u,\mathbf v)=\operatorname{mod}\left[\operatorname{atan2}(u_xv_y-u_yv_x,\mathbf u\cdot\mathbf v),2\pi\right].
\end{equation}
Periodic cycles are unwrapped with minimum-image displacements before geometric analysis. The projected area of a ring $\ell$ is obtained from the shoelace formula,
\begin{equation}
A_\ell=\frac{1}{2}\left|\sum_{t=1}^{R_\ell}(x_ty_{t+1}-y_tx_{t+1})\right|.
\label{eq:area}
\end{equation}
Because Eq.~\ref{eq:area} is an in-plane projected area, it is meaningful in the present form only for planar or nearly planar sheets with an unambiguous planar embedding. The vacancy structures used below were relaxed with the out-of-plane coordinate constrained, so this condition is satisfied by construction.

\subsection{Ring observables and the modeling vector}
For ring size $R$, the face fraction is
\begin{equation}
p_R=\frac{N_R}{N_{\rm faces}},
\end{equation}
and the ring-size Shannon entropy is\cite{Shannon1948}
\begin{equation}
S_{\rm topo}=-\sum_Rp_R\ln p_R.
\label{eq:entropy}
\end{equation}
A ring topological charge is defined relative to graphene as
\begin{equation}
q_\ell=6-R_\ell,
\label{eq:qdef}
\end{equation}
from which we calculate $\langle q\rangle$, $\langle|q|\rangle$, and $\operatorname{Var}(q)$. Ring geometry is summarized by the mean, median, standard deviation, coefficient of variation, 25th and 75th percentiles, maximum projected area, and the fraction of total face area associated with $R\ge10$. A dual graph connects rings sharing an atomic edge, allowing the fractions of 5--7 adjacencies and non-hexagon--non-hexagon adjacencies to be calculated.

The 19 variables used as machine-learning inputs are listed explicitly in Table~\ref{tab:features}. Two additional quantities generated by the implementation, $N_{\rm faces}$ and $d_{\max}$, are retained as diagnostics but excluded from machine learning: $N_{\rm faces}$ can encode analysis-cell size, and $d_{\max}=1.40d_0$ is algebraically redundant with $d_0$.

\begin{table}[t]
\caption{Core TGD variables used for machine learning. Areas are projected in-plane areas.}
\label{tab:features}
\begin{tabular}{lll}
\toprule
Category & Components & Units \\
\midrule
Ring fractions & $p_5,p_6,p_7,p_{8+}$ & dimensionless \\
Ring diversity & $S_{\rm topo}$ & dimensionless \\
Topological charge & $\langle q\rangle,\langle|q|\rangle,\mathrm{Var}(q)$ & dimensionless \\
Ring geometry & $\bar A,A_{\rm med},\sigma_A,\mathrm{CV}_A,A_{25},A_{75},A_{\max},f_{A,R\ge10}$ & \AA$^2$ or dimensionless \\
Dual graph & $f_{5-7},f_{\rm defect-defect}$ & dimensionless \\
Bonding scale & $d_0$ & \AA \\
\bottomrule
\end{tabular}
\end{table}

\subsection{Datasets and target properties}
\paragraph{Published 120-allotrope dataset.}
The first dataset contains 120 published planar carbon allotropes from Shi \textit{et al.}\cite{Shi2021} We use the energetic quantity supplied with that dataset and relabel it explicitly as a relative energy per atom,
\begin{equation}
\Delta E_i=\frac{E_i}{N_i}-\frac{E_{\rm graphene}}{N_{\rm graphene}},
\label{eq:deltaE}
\end{equation}
reported in eV atom$^{-1}$, with graphene defining zero. These electronic-structure values were imported from the published dataset; they were not recomputed in this work. The structures are used here to test whether the descriptor can learn variation in the published $\Delta E$ values. This definition replaces the ambiguous ``total energy'' terminology used in the earlier manuscript version.

\paragraph{Vacancy-defected graphene dataset.}
The second dataset contains 360 independently generated vacancy configurations derived from a 300-atom pristine graphene sheet. Twelve vacancy concentrations, 2--24\% in 2\% steps, are represented by 30 realizations each. Defected cells therefore contain 228--294 carbon atoms before descriptor replication. The archived calculation scripts use DFTB+ through ASE with the GFN1-xTB Hamiltonian,\cite{dftb,Grimme2017GFN1} $\Gamma$-point sampling, and BFGS structural relaxation with the cell fixed. The $z$ coordinate of every atom is constrained while $x$ and $y$ are relaxed, using a maximum force criterion of $0.02$ eV \AA$^{-1}$ and at most 200 BFGS steps. The target used for all vacancy regressions is the formation energy per vacancy,
\begin{equation}
E_f^{\rm vac}=\frac{E_{\rm def}-(N_{\rm def}/N_0)E_0}{N_{\rm vac}},
\label{eq:vac}
\end{equation}
in eV vacancy$^{-1}$. Here $E_0$ and $N_0$ refer to pristine graphene evaluated with the same method, and $N_{\rm vac}=N_0-N_{\rm def}$. Total energy per atom and relaxation-derived quantities are \emph{not} included among the input features in the revised vacancy analysis.

\subsection{Statistical validation and baselines}
For the 120-allotrope dataset, we compare Ridge regression, support-vector regression, random forests, and gradient boosting.\cite{HoerlKennard1970,Drucker1997,randomforest,Friedman2001} Reported means and standard deviations are obtained from repeated five-fold cross-validation with 20 repeats (100 validation folds). A separate shuffled 10-fold partition is used only to generate out-of-fold points for visualization.

For the vacancy dataset, random splitting would place configurations with the same vacancy concentration in both train and test sets. We therefore group by concentration and use leave-one-concentration-out validation (12 folds). The principal control experiment uses the same Ridge estimator for four input sets: vacancy concentration alone, non-hexagon fraction $1-p_6$ alone, $\langle|q|\rangle$ alone, and the 19-component TGD. Keeping the estimator fixed makes the comparison a test of input information rather than model flexibility. Uncertainty is additionally assessed with 100 group-shuffle splits in which whole concentrations remain together.

For the TGD--Matminer comparison, both representations are evaluated with the same gradient-boosting regressor and identical repeated five-fold partitions (10 repeats). The Matminer matrix contains 203 numerical input features after removing the target and atom count, whereas the TGD matrix contains 19.

\section{Results and Discussion}
\subsection{Physical interpretability of the descriptor}
\begin{figure}[t!]
\centering
\includegraphics[width=0.9\linewidth]{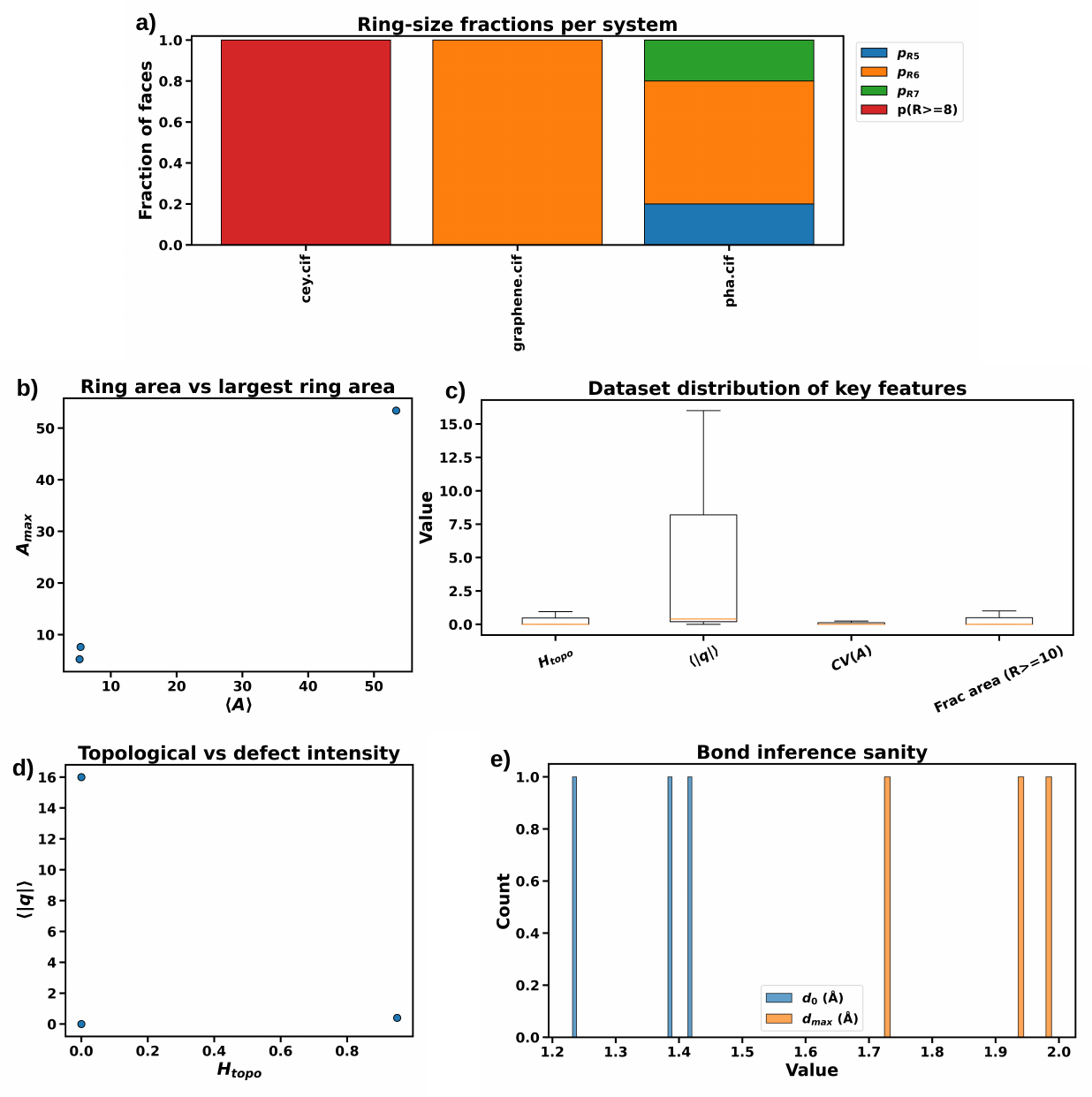}
\caption{Interpretability of TGD for representative planar carbon networks: graphene, phagraphene, and cey-graphene. Ring fractions, projected ring geometry, topological entropy, mean absolute topological charge, and the adaptive bond scale distinguish hexagonal, mixed 5--6--7, and porous large-ring regimes.}
\label{fig:fig1}
\end{figure}

Figure~\ref{fig:fig1} illustrates the descriptor on graphene, phagraphene, and cey-graphene, three systems with qualitatively different ring motifs. Graphene gives $p_6=1$, $S_{\rm topo}=0$, and $\langle|q|\rangle=0$. Phagraphene produces the expected 5--6--7 mixture and nearly neutral mean topological charge. Cey-graphene is dominated by large rings and has a much larger projected ring area. The purpose of these examples is interpretive: the coordinates in descriptor space correspond directly to observable graph motifs rather than latent variables.

The descriptor also separates ring diversity from deviation from hexagonal order. A network composed mostly of one large ring class can have low $S_{\rm topo}$ but high $\langle|q|\rangle$, whereas a 5--6--7 network may have higher entropy but a smaller mean absolute charge. This distinction is useful because no single scalar is assumed to represent all forms of disorder.

\subsection{Predictive accuracy for the 120-allotrope dataset}
\begin{figure}[t!]
\centering
\includegraphics[width=0.92\linewidth]{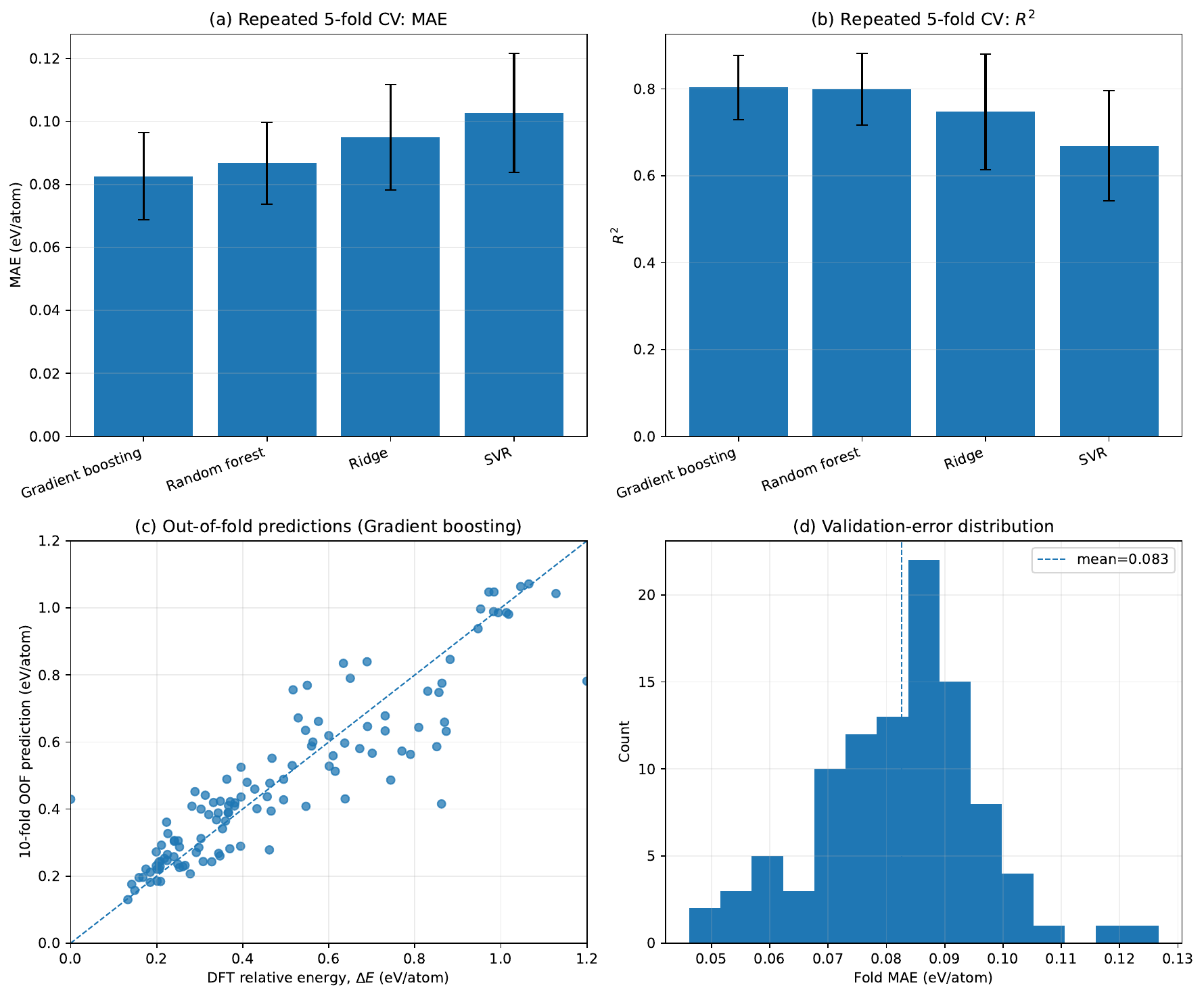}
\caption{Predictive validation for the 120-allotrope target $\Delta E$ (eV atom$^{-1}$). (a,b) Mean and standard deviation of MAE and $R^2$ across repeated five-fold cross-validation (20 repeats). (c) Ten-fold out-of-fold predictions for the best mean-CV model. (d) Distribution of fold MAEs.}
\label{fig:fig2}
\end{figure}

Figure~\ref{fig:fig2} reports repeated-cross-validation results for $\Delta E$ rather than training-set fits. Gradient boosting gives the lowest mean MAE, $0.0826\pm0.0138$ eV atom$^{-1}$, with $R^2=0.803\pm0.074$. Random forests give $0.0867\pm0.0130$ eV atom$^{-1}$ and $R^2=0.799\pm0.083$; Ridge and SVR are less accurate under the same partitions. The independent 10-fold out-of-fold visualization for gradient boosting gives an aggregate MAE of 0.0808 eV atom$^{-1}$ and $R^2=0.817$. These values quantify predictive performance without claiming absence of overfitting or transferability beyond the sampled structural domain.

\begin{figure}[t!]
\centering
\includegraphics[width=0.9\linewidth]{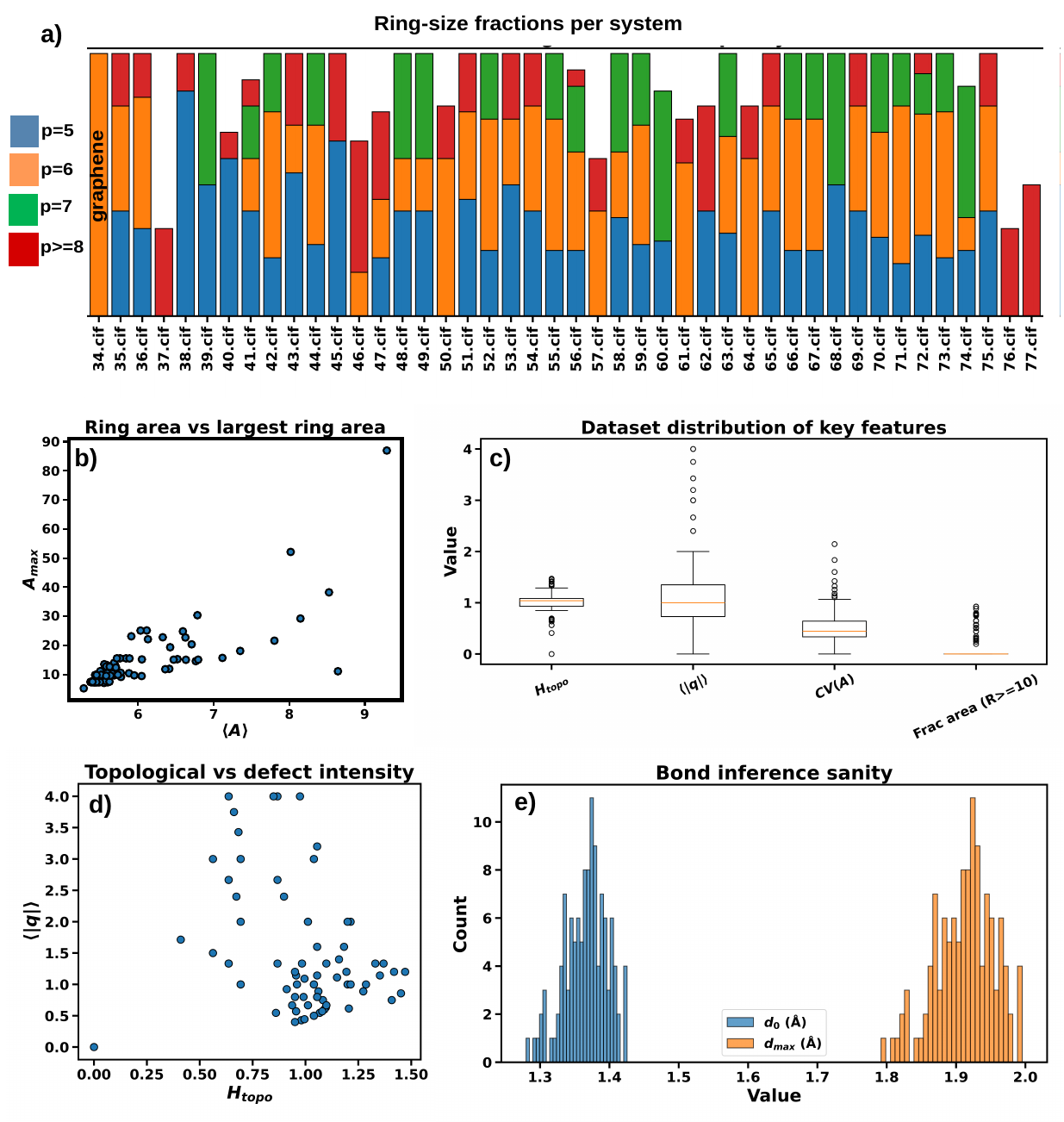}
\caption{Descriptor-space interpretation across the 120 planar carbon allotropes. The panels summarize ring fractions, projected ring areas, distributions of selected TGD variables, entropy--charge relationships, and adaptive bond scales.}
\label{fig:fig3}
\end{figure}

Figure~\ref{fig:fig3} is complementary to Fig.~\ref{fig:fig2}: it addresses interpretation rather than accuracy. Across the 120 structures, ring fractions, area statistics, entropy, and topological-charge moments cover distinct structural regimes. The observed correlations are descriptive and do not by themselves establish causality; they show which graph-level motifs the model has access to.

\subsection{Vacancy energetics: grouped validation and trivial controls}
\begin{figure}[t!]
\centering
\includegraphics[width=0.92\linewidth]{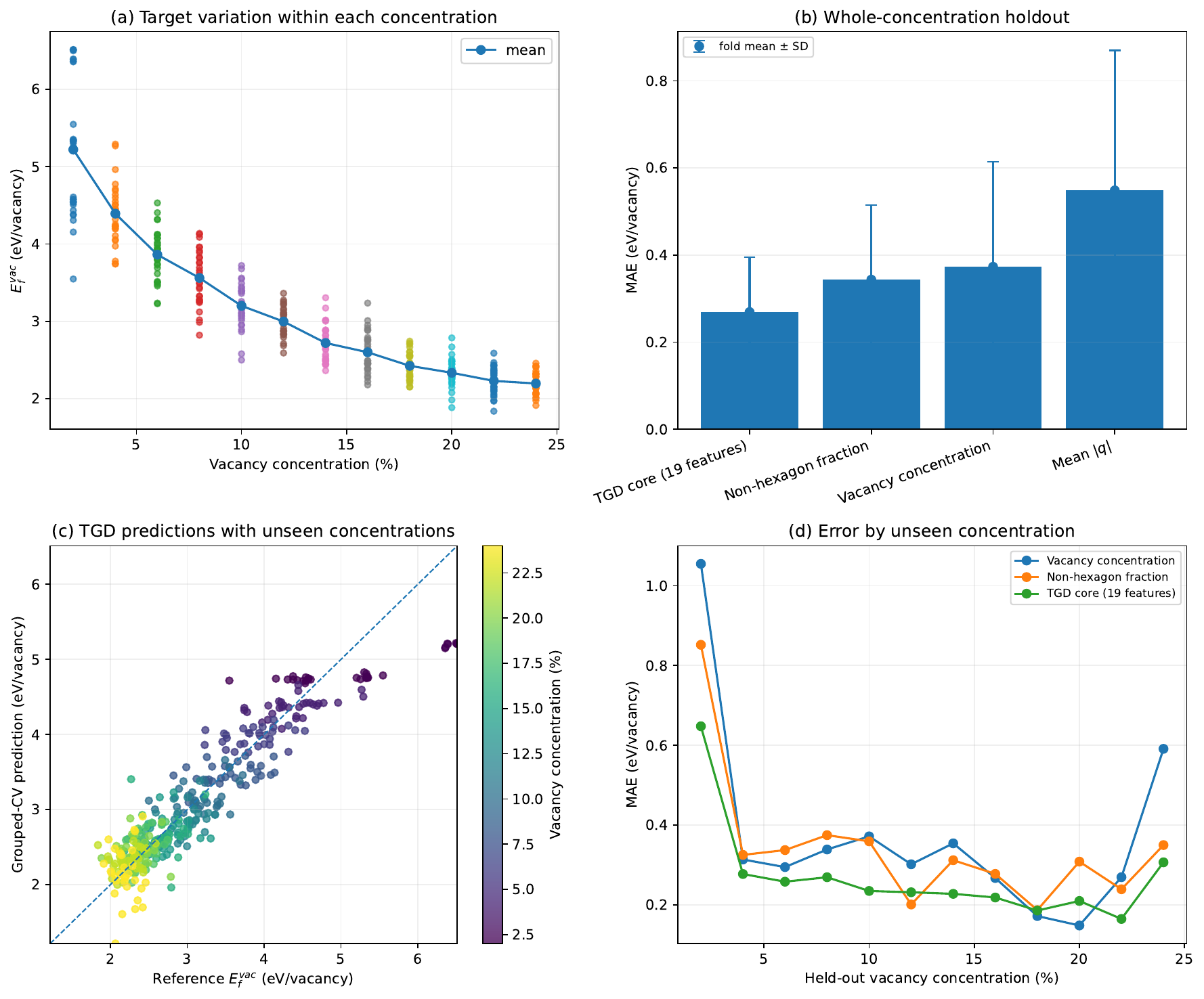}
\caption{Group-aware validation for vacancy formation energy per vacancy, $E_f^{\rm vac}$ (eV vacancy$^{-1}$). (a) Target distribution at 2--24\% vacancy concentration. (b) Leave-one-concentration-out MAE for vacancy concentration, non-hexagon fraction, $\langle|q|\rangle$, and TGD. (c) TGD predictions when entire concentrations are unseen during training. (d) MAE as a function of the held-out concentration. All control models use the same Ridge estimator.}
\label{fig:fig4}
\end{figure}

The vacancy dataset presents a stricter test because vacancy concentration simultaneously influences energy and global ring statistics. Figure~\ref{fig:fig4}(a) shows that $E_f^{\rm vac}$ changes systematically with concentration but retains substantial configuration-to-configuration spread within each concentration. Holding out entire concentrations, concentration alone gives a pooled MAE of 0.373 eV vacancy$^{-1}$ ($R^2=0.705$), while non-hexagon fraction alone gives 0.343 eV vacancy$^{-1}$ ($R^2=0.757$). The one-variable $\langle|q|\rangle$ control is weaker, with a MAE of 0.548 eV vacancy$^{-1}$ ($R^2=0.417$).

The 19-component TGD reduces the pooled MAE to 0.269 eV vacancy$^{-1}$ and raises pooled $R^2$ to 0.861. Its mean leave-one-concentration-out MAE is $0.269\pm0.125$ eV vacancy$^{-1}$. Across 100 group-shuffle splits, the TGD MAE is $0.275\pm0.063$ eV vacancy$^{-1}$. Thus defect count is an important baseline, but it does not account for all predictive information present in the revised TGD input. We emphasize that this is a comparative information-content test within the present vacancy dataset, not evidence of universal transferability to arbitrary defect chemistries.

\begin{figure}[t!]
\centering
\includegraphics[width=0.9\linewidth]{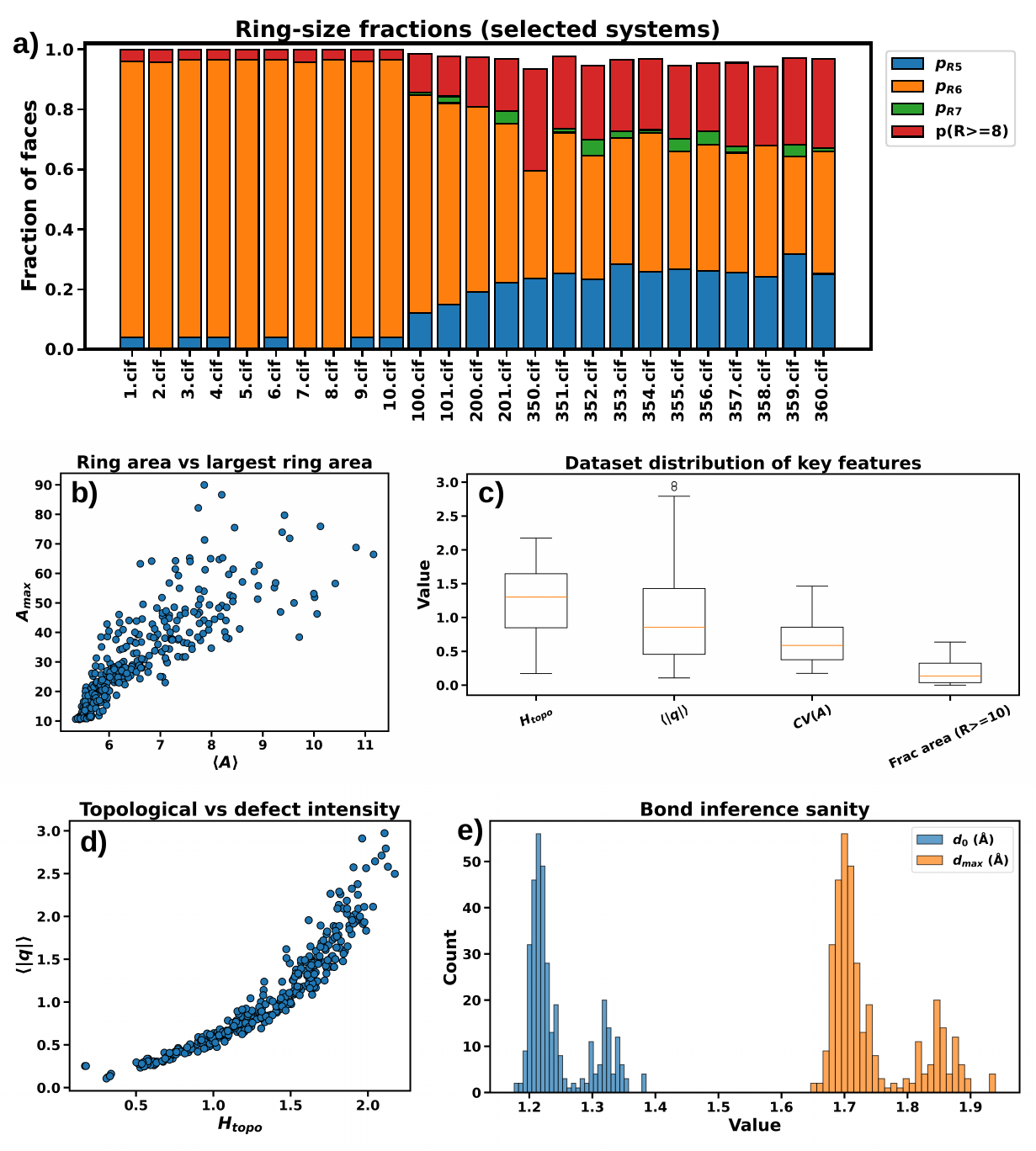}
\caption{Interpretability analysis for the 360 vacancy-defected graphene configurations spanning 2--24\% vacancy concentration. The panels show changes in ring populations, projected ring areas, selected TGD distributions, entropy--charge relationships, and inferred bonding scales.}
\label{fig:fig5}
\end{figure}

Figure~\ref{fig:fig5} shows the structural interpretation of this vacancy series. Hexagonal-ring fraction decreases as vacancies accumulate, while non-hexagonal and large-ring content grows. Area heterogeneity and topological-charge statistics likewise evolve with defect loading. Importantly, configurations at the same concentration need not share identical ring arrangements. This is precisely the regime in which global concentration and global non-hexagon fraction become incomplete descriptors.

\subsection{TGD versus Matminer under identical cross-validation}
\begin{figure}[t!]
\centering
\includegraphics[width=0.92\linewidth]{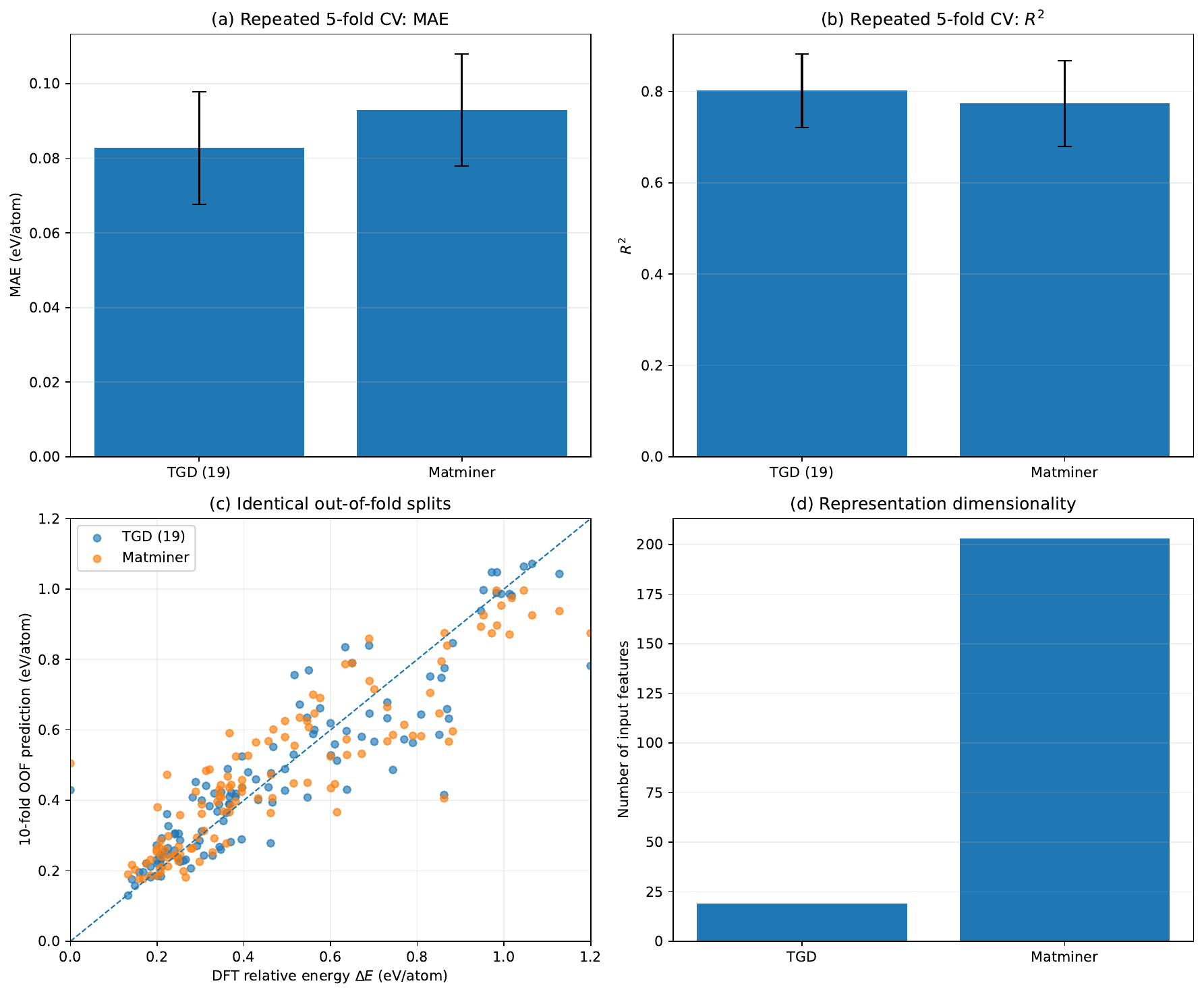}
\caption{Controlled TGD--Matminer comparison for the 120-allotrope $\Delta E$ target. (a,b) Repeated five-fold CV MAE and $R^2$ under the same gradient-boosting model. (c) Out-of-fold predictions using identical folds. (d) Number of input features: 19 for TGD and 203 for Matminer.}
\label{fig:fig6}
\end{figure}

Figure~\ref{fig:fig6} compares the two representations using the same gradient-boosting model and identical repeated folds. TGD uses 19 model inputs, whereas the Matminer matrix contains 203. TGD gives a repeated-CV MAE of $0.0828\pm0.0151$ eV atom$^{-1}$ and $R^2=0.802\pm0.080$; Matminer gives $0.0929\pm0.0150$ eV atom$^{-1}$ and $R^2=0.774\pm0.094$. We therefore describe the performances as comparable within this dataset and protocol, rather than making a general state-of-the-art claim. The dimensionality difference is substantial, but it should not be interpreted as a universal ranking of descriptor families.

The interpretability advantage is conceptually separate from the accuracy comparison. A full-data gradient-boosting fit assigns its largest TGD importances to $\langle|q|\rangle$, mean projected ring area, $d_0$, and area quantiles. These quantities can be inspected directly as structural observables. Feature importance, however, is model-dependent and correlated features can share or exchange importance; we therefore use it as an explanatory aid, not as proof of a unique microscopic mechanism.

\subsection{One-variable analytical surrogate}
Because $\langle|q|\rangle$ is a prominent TGD variable for the 120-allotrope dataset, we test whether it supports a useful one-dimensional empirical surrogate. To remove the scale non-identifiability of a four-parameter rational function, we fit
\begin{equation}
\Delta E=a+\frac{b\,\langle|q|\rangle}{1+c\,\langle|q|\rangle}.
\label{eq:surrogate}
\end{equation}
The full-data estimates are $a=0.0698$, $b=0.4050$, and $c=0.1499$ for $\Delta E$ in eV atom$^{-1}$. Over repeated five-fold validation, the rational surrogate gives MAE $0.1191\pm0.0170$ eV atom$^{-1}$ and mean $R^2=0.650\pm0.158$; a linear model gives $0.1208\pm0.0176$ eV atom$^{-1}$ and $R^2=0.645\pm0.147$. The improvement from the rational form is therefore small.

\begin{figure}[t!]
\centering
\includegraphics[width=0.98\linewidth]{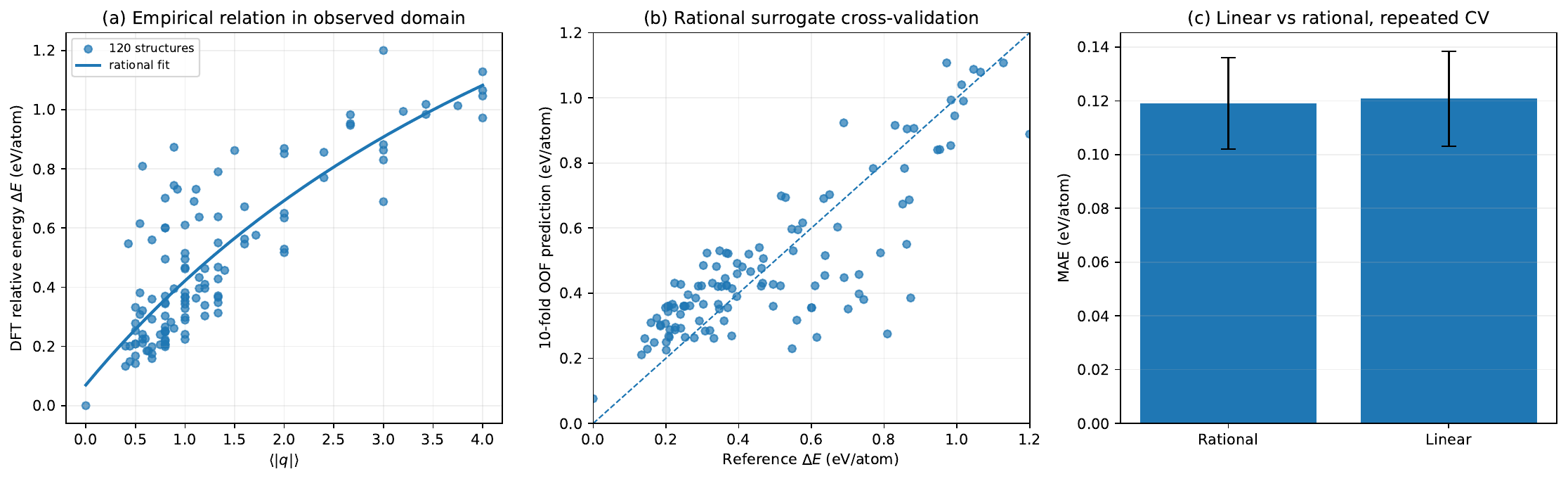}
\caption{Dataset-specific one-variable surrogate based on $\langle|q|\rangle$ for the 120-allotrope relative-energy target. (a) Rational fit over the observed domain, $0\le\langle|q|\rangle\le4$. (b) Ten-fold out-of-fold predictions. (c) Repeated-CV comparison with a linear model.}
\label{fig:fig7}
\end{figure}

Equation~\ref{eq:surrogate} should be interpreted as a compact surrogate for this dataset, not as a universal carbon-energy law. In particular, its finite asymptote as $\langle|q|\rangle\rightarrow\infty$ has not been assigned physical meaning, and the fit has not been validated outside the observed range $0\le\langle|q|\rangle\le4$. It is also distinct from the vacancy analysis: Fig.~\ref{fig:fig4} shows that $\langle|q|\rangle$ alone is insufficient to match the full descriptor for $E_f^{\rm vac}$. This separation resolves the ambiguity between the two targets in the earlier manuscript version.

\subsection{Supercell convergence and computational scaling}
\begin{figure}[t!]
\centering
\includegraphics[width=0.92\linewidth]{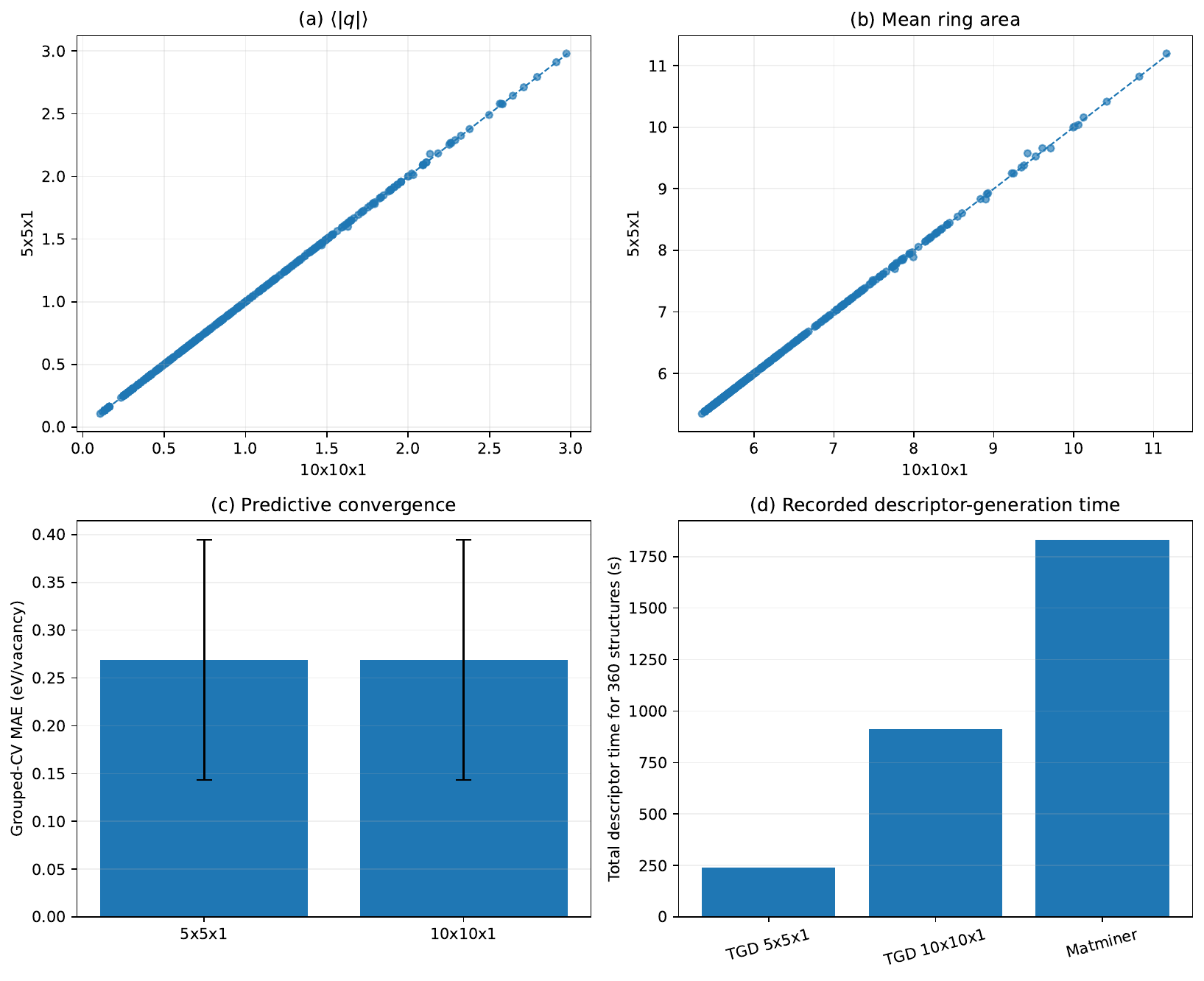}
\caption{Supercell convergence for the vacancy dataset. (a,b) Agreement of representative TGD variables between $5\times5\times1$ and $10\times10\times1$ replications. (c) Leave-one-concentration-out MAE using the same 19 TGD inputs. (d) Recorded descriptor-generation times for the 360-structure workflow.}
\label{fig:fig8}
\end{figure}

For large vacancy structures, descriptor statistics converge with substantially less replication than was used as the conservative default. Figure~\ref{fig:fig8} compares $5\times5\times1$ and $10\times10\times1$ replications. Representative features such as $\langle|q|\rangle$ and mean ring area lie close to the identity line, and the 19-component grouped-CV result is nearly unchanged: pooled MAEs are 0.2690 and 0.2692 eV vacancy$^{-1}$, respectively. The recorded descriptor-generation times for the full set of 360 structures were 238.20 s for TGD at $5\times5\times1$ and 911.49 s at $10\times10\times1$, compared with 1831.89 s for the Matminer workflow used in the timing comparison. These timings are implementation- and hardware-dependent and are reported only for the present workflow.

\subsection{Robustness of the adaptive bonding criterion}
\begin{figure}[t!]
\centering
\includegraphics[width=0.98\linewidth]{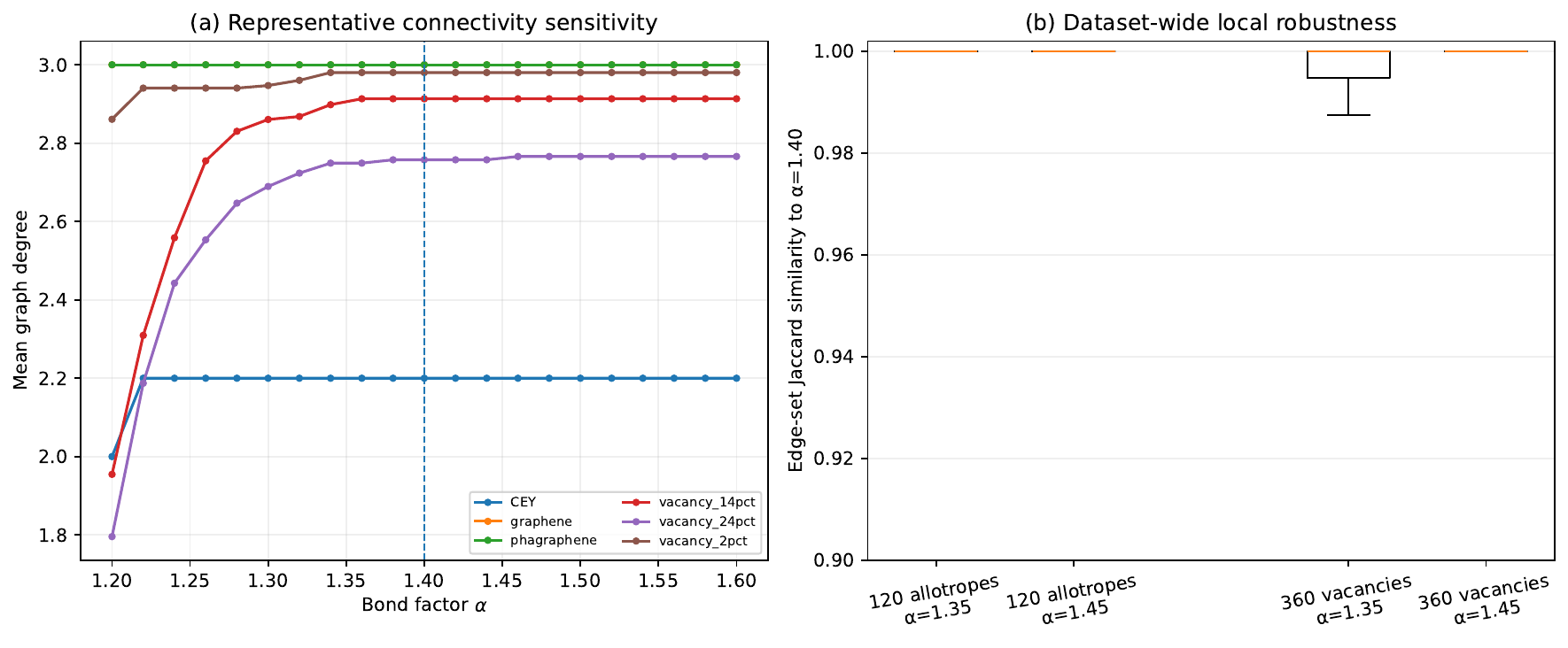}
\caption{Sensitivity of inferred connectivity to the adaptive bond factor. (a) Mean graph degree as $\alpha$ is varied for representative planar structures. (b) Dataset-wide edge-set Jaccard similarity at $\alpha=1.35$ and 1.45 relative to $\alpha=1.40$.}
\label{fig:fig9}
\end{figure}

Reviewer concerns about anchoring the cutoff to the shortest bond motivate an explicit sensitivity analysis. Figure~\ref{fig:fig9} scans $\alpha=1.20$--1.60 for representative structures and compares edge sets at $\alpha=1.35$ and 1.45 against the production value 1.40 for every structure. For all 120 allotropes, the median and fifth-percentile edge-set Jaccard similarities are 1.000 for both perturbations. For the 360 vacancy structures, the median remains 1.000; the fifth percentiles are 0.989 at $\alpha=1.35$ and 0.997 at $\alpha=1.45$.

The mixed-$sp/sp^2$ cey-graphene example is informative because $d_0=1.232$~\AA\ while the longest accepted bond at $\alpha=1.40$ is 1.486~\AA. Its graph contains both degree-2 and degree-3 sites rather than forcing threefold coordination. The most heavily defected representative (24\%) has a small fraction (0.85\%) of sites outside degree 2 or 3, illustrating the regime in which local reconstruction begins to challenge a single global threshold. Accordingly, the adaptive rule is robust for the present datasets but should not be assumed reliable without checking when a structure contains extreme local strain, unusually short bonds, heteroatomic bond classes, or out-of-plane reconstructions.

\subsection{Current limitations and intended applicability}
TGD is deliberately specialized. First, the face-tracing and area construction assumes a single planar embedding. Buckled sheets may be approximately projected only when that projection remains one-to-one; strongly curved membranes, nanotubes, multilayers, and crossed networks require a different surface-aware definition. Second, the current carbon-specific global cutoff and the charge convention $q=6-R$ are not directly transferable to chemically heterogeneous 2D materials. Heteroatomic systems would require species-dependent bonding criteria and a reconsideration of which reference ring/coordination carries physical meaning.

Third, global ring statistics are not injective: distinct local arrangements can share the same ring histogram, charge moments, or area distribution. The dual-graph adjacency terms partially encode spatial organization, but they do not capture all defect separations, orientations, elastic fields, or reconstruction motifs. This limitation is visible in the spread of vacancy energies at fixed concentration and motivates future local or distance-resolved correlation features. Finally, Eq.~\ref{eq:surrogate} is only a dataset-level approximation. Its coefficients and asymptote should be refitted and independently validated before use on another structural family.

\section{Data Availability}
The data supporting this article comprise the published dataset of 120 planar carbon allotropes cited in the manuscript and the vacancy-defected graphene dataset generated in this study. The data generated in this study, including processed descriptor tables and analysis outputs, are available from the corresponding author upon reasonable request.

\section{Code Availability}
The TGD implementation and representative structural files are openly available at \url{https://github.com/tromer-unb/TGD}. Additional analysis scripts supporting the results reported here are available from the corresponding author upon reasonable request.

\section{Summary and conclusions}
We presented TGD as a compact, interpretable representation for planar two-dimensional carbon networks and revised the validation to make the target quantities, units, scope, and statistical controls explicit. The 120-allotrope task predicts the DFT relative energy per atom, $\Delta E$ (eV atom$^{-1}$), whereas the vacancy task predicts the DFTB+/GFN1-xTB vacancy formation energy per vacancy, $E_f^{\rm vac}$ (eV vacancy$^{-1}$). These quantities are no longer conflated.

Repeated cross-validation on the 120 structures shows that a 19-component TGD supports predictive errors comparable to the tested 203-component Matminer representation under the same gradient-boosting protocol. In the vacancy dataset, holding out complete concentrations and comparing against simple controls shows that vacancy concentration and non-hexagon fraction explain substantial variation, as expected, but that the full TGD lowers the pooled MAE further. This result supports additional information content in the combined topology--geometry representation without implying that defect count is unimportant.

The interpretive contribution is separate: ring fractions, charge moments, area statistics, and dual-graph terms retain a direct structural meaning. The adaptive bond criterion is stable to modest cutoff perturbations over the present datasets, including mixed $sp/sp^2$ connectivity, but its limitations under extreme reconstruction are now stated explicitly. The one-variable $\langle|q|\rangle$ expression is retained only as a dataset-specific surrogate over the sampled domain and is not presented as a universal physical law. Within these boundaries, TGD provides a computationally compact framework for comparing planar carbon networks while preserving a transparent connection between graph topology, ring geometry, and learned energetic trends.

\section*{Acknowledgements}
RMT acknowledges CNPq for financial support (grants 307371/2025-5 and 444069/2024-0).

\section*{CREDIT AUTHORSHIP STATEMENT}
R\^omulo Uriel Liberal de Oliveira -- Data curation; Formal analysis; Investigation; Methodology; Software; Validation; Visualization; Writing -- original draft.

Raphael Matozo Tromer -- Conceptualization; Formal analysis; Investigation; Methodology; Project administration; Resources; Supervision; Validation; Writing -- original draft; Writing -- review and editing.

\bibliographystyle{unsrt}
\bibliography{example}

\end{document}